\documentclass[pra,amsmath,showpacs,twocolumn,superscriptaddress,longbibliography]{revtex4-1}

\usepackage{graphicx}
\usepackage{units}
\usepackage{color}
\usepackage{float}

\begin{document}
\title{Emergence of chaotic scattering in ultracold Er and Dy}

\author{T. Maier}
\affiliation{5.~Physikalisches Institut and Center for Integrated Quantum Science and Technology,
Universit\"at Stuttgart, Pfaffenwaldring 57, 70550 Stuttgart, Germany}
\author{H. Kadau}
\affiliation{5.~Physikalisches Institut and Center for Integrated Quantum Science and Technology,
Universit\"at Stuttgart, Pfaffenwaldring 57, 70550 Stuttgart, Germany}
\author{M. Schmitt}
\affiliation{5.~Physikalisches Institut and Center for Integrated Quantum Science and Technology,
Universit\"at Stuttgart, Pfaffenwaldring 57, 70550 Stuttgart, Germany}
\author{M. Wenzel}
\affiliation{5.~Physikalisches Institut and Center for Integrated Quantum Science and Technology,
Universit\"at Stuttgart, Pfaffenwaldring 57, 70550 Stuttgart, Germany}
\author{I. Ferrier-Barbut}
\affiliation{5.~Physikalisches Institut and Center for Integrated Quantum Science and Technology,
Universit\"at Stuttgart, Pfaffenwaldring 57, 70550 Stuttgart, Germany}
\author{T. Pfau}
\affiliation{5.~Physikalisches Institut and Center for Integrated Quantum Science and Technology,
Universit\"at Stuttgart, Pfaffenwaldring 57, 70550 Stuttgart, Germany}
\author{A. Frisch}
\affiliation{Institut f\"ur Experimentalphysik, Universit\"at Innsbruck, Technikerstra{\ss}e 25, 6020 Innsbruck, Austria}
\affiliation{Institut f\"ur Quantenoptik und Quanteninformation, \"Osterreichische Akademie der Wissenschaften, 6020 Innsbruck, Austria}
\author{S. Baier}
\affiliation{Institut f\"ur Experimentalphysik, Universit\"at Innsbruck, Technikerstra{\ss}e 25, 6020 Innsbruck, Austria}
\author{K. Aikawa}
\altaffiliation{Current adress: Department of Physics, Graduate School of Science and Engineering, Tokyo Institute of Technology, Meguro-ku, Tokyo, 152-8550 Japan.}
\affiliation{Institut f\"ur Experimentalphysik, Universit\"at Innsbruck, Technikerstra{\ss}e 25, 6020 Innsbruck, Austria}
\author{L. Chomaz}
\affiliation{Institut f\"ur Experimentalphysik, Universit\"at Innsbruck, Technikerstra{\ss}e 25, 6020 Innsbruck, Austria}
\affiliation{Institut f\"ur Quantenoptik und Quanteninformation, \"Osterreichische Akademie der Wissenschaften, 6020 Innsbruck, Austria}
\author{M. J. Mark}
\affiliation{Institut f\"ur Experimentalphysik, Universit\"at Innsbruck, Technikerstra{\ss}e 25, 6020 Innsbruck, Austria}
\affiliation{Institut f\"ur Quantenoptik und Quanteninformation, \"Osterreichische Akademie der Wissenschaften, 6020 Innsbruck, Austria}
\author{F. Ferlaino}
\affiliation{Institut f\"ur Experimentalphysik, Universit\"at Innsbruck, Technikerstra{\ss}e 25, 6020 Innsbruck, Austria}
\affiliation{Institut f\"ur Quantenoptik und Quanteninformation, \"Osterreichische Akademie der Wissenschaften, 6020 Innsbruck, Austria}
\author{C. Makrides}
\affiliation{Department of Physics, Temple University, Philadelphia, Pennsylvania 19122, USA}
\author{E. Tiesinga}
\affiliation{Joint Quantum Institute and Center for Quantum Information and Computer Science,
National Institute of Standards and Technology and the University of Maryland, 100 Bureau Drive, Stop 8423, Gaithersburg, Maryland 20899, USA}
\author{A. Petrov}
\altaffiliation{Alternative address: NRC ``Kurchatov Institute'' PNPI 188300; Division of Quantum Mechanics, St. Petersburg State University, 198904, Russia.}
\author{S. Kotochigova}
\affiliation{Department of Physics, Temple University, Philadelphia, Pennsylvania 19122, USA}

\date{\today}

\pacs{03.65.Nk, 34.50.-s, 05.45.Mt}

\begin{abstract}
We show that for ultracold magnetic lanthanide atoms chaotic scattering
emerges due to a combination of anisotropic interaction potentials and
Zeeman coupling under an external magnetic field. This scattering is
studied in a collaborative experimental and theoretical effort for both
dysprosium and erbium. We present extensive atom-loss measurements of
their dense magnetic Feshbach resonance spectra, analyze their statistical
properties, and compare to predictions from a random-matrix-theory
inspired model.  Furthermore, theoretical coupled-channels simulations
of the anisotropic molecular Hamiltonian at zero magnetic field show that
weakly-bound, near threshold diatomic levels form overlapping, uncoupled
chaotic series that when combined are randomly distributed. The Zeeman
interaction shifts and couples these levels,  leading to a Feshbach
spectrum of zero-energy bound states with nearest-neighbor spacings
that changes from randomly to chaotically distributed for increasing
magnetic field.  Finally, we show that the extreme temperature sensitivity
of a small, but sizeable fraction of the resonances in the Dy and Er
atom-loss spectra is due to resonant non-zero partial-wave collisions.
Our threshold analysis for these resonances indicates a large collision-energy
dependence of the three-body recombination rate.  \end{abstract}

\maketitle


\section{Introduction}\label{intro}

Anisotropic interactions are a central and modern tool for engineering
quantum few- and many-body processes \cite{CMatterDDI}.  A prominent
example of such an interaction is the long-range dipole-dipole
interaction (DDI) acting for instance between polar molecules
\cite{PolMolRev}, Rydberg atoms \cite{RydbergRev}, or  magnetic atoms
\cite{DDIRev}. Over the years, fascinating quantum effects of the
anisotropy have been observed, such as the $d$-wave collapse of a
dipolar Bose-Einstein condensate \cite{dwave}, the deformation of the
Fermi surface \cite{Fermisurface}, and the control of  stereodynamics
in dipolar collisions \cite{stereodynamics}. Moreover, the DDI is
expected to give rise to a plethora of  few- and many-body phenomena,
which still await  observation, such as universal few-body physics
\cite{quasiuniv,3BPhysics}, rotonic features \cite{Roton, Roton2},
two-dimensional stable solitons \cite{2DSoliton}, and the supersolid
phase \cite{Supersolid}.

Recently, atomic species in the lanthanide family became available
to the field of ultracold quantum gases. The interaction between
magnetic lanthanide atoms, such as Er \cite{ErBEC, ErFermion} and Dy
\cite{DyBEC, DyFermion}, is highly anisotropic. This is  not only due
to the long-range DDI, originating from their large magnetic moment,
but also to the shorter-ranged van-der-Waals interaction \cite{ADIFR},
which exhibits anisotropic contributions arising from the large orbital
angular momentum of their valence electrons.

For magnetic lanthanides, which also include the successfully laser-cooled
elements Ho \cite{HoMOT} and Tm \cite{TmMOT}, the orbital anisotropy
is a consequence of a partially filled submerged 4f electron shell
that underlies a closed outer 6s shell. This leads to an electronic
ground state with a  total atomic angular momentum $\vec \jmath$ with
$j\gg1$. Consequently, in collisions between such atoms there exist
$(j+1)^2$ non-degenerate ({\it gerade}) molecular Born-Oppenheimer (BO)
potentials and a correspondingly large manifold of collision channels
with associated molecular bound states.  This is in sharp contrast
to the one or two BO potentials encountered in alkaline-earth
and alkali-metal atom collisions. In addition, the anisotropy or
orientation dependence of the BO potentials strongly mixes collision
channels with large relative orbital angular momentum $\vec \ell$
between the atoms even for our ultracold collisions with a  $\ell=0$,
$s$-wave initial channel. The complexity of the molecular forces are
reflected in a dense spectrum of Fano-Feshbach resonances as a function
of magnetic field $B$, as recently observed in Er \cite{ErBEC,ErstatFR} and
Dy \cite{DyFeshbach}. In Er a statistical analysis of the spacings
between  resonances  has shown correlations that revealed chaotic
scattering. The data set of the initial Dy experiments was too small to
extract statistically-significant correlations.

\begin{figure*}
\includegraphics[width=2\columnwidth]{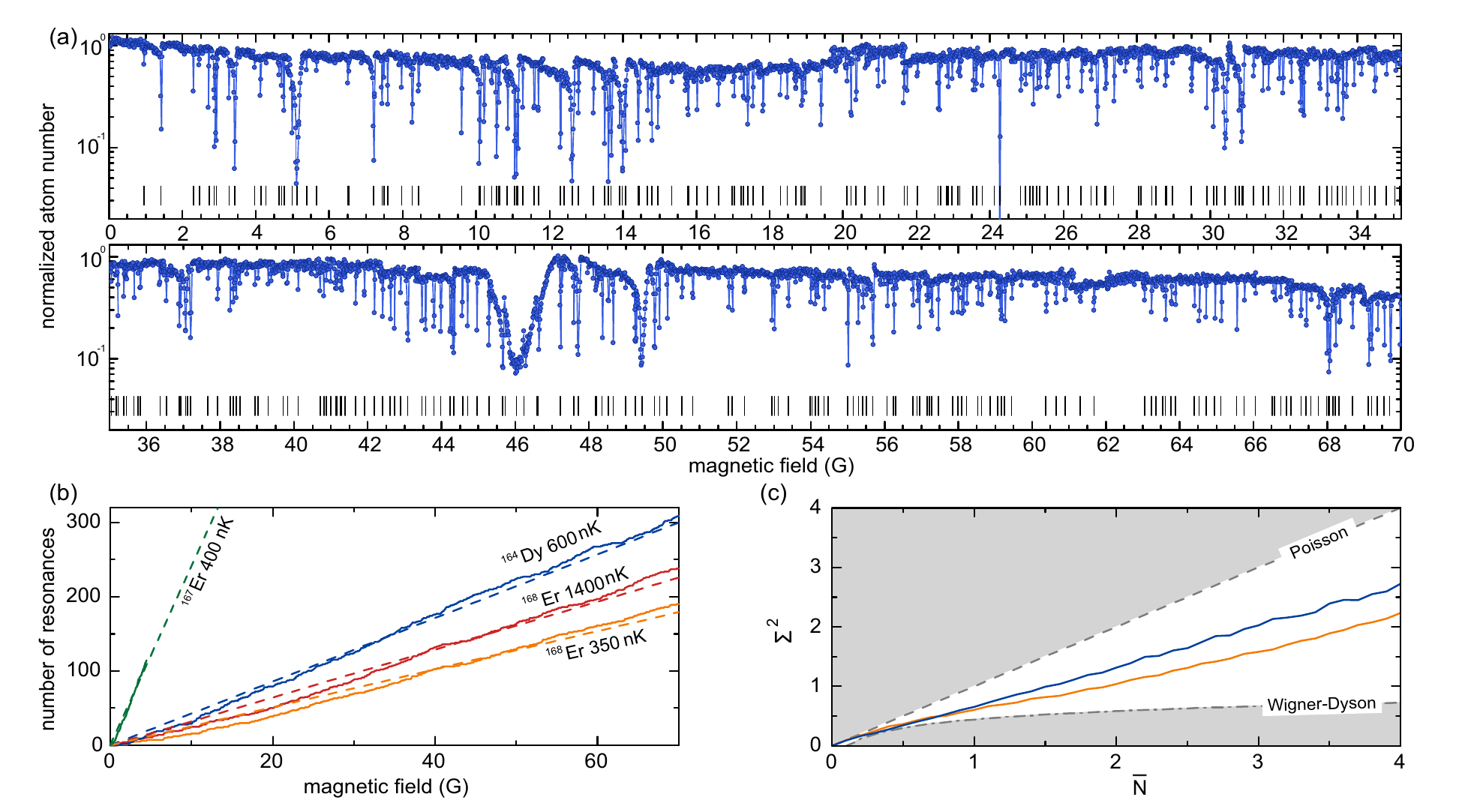}
\caption{(color online) (a) Trap-loss spectroscopy mapping of the
Fano-Feshbach spectrum of $^{164}$Dy as a function of magnetic field
between $B=$0$\,$G to 70$\,$G with a data point every 14.5$\,$mG and temperature $T=600$ nK. Each
data point is an average of three measurements. (b) Staircase function
for the number of resonances as a function of $B$ for $^{164}$Dy,
$^{168}$Er at two temperatures, and fermionic $^{167}$Er. Dashed
lines are linear fits forced to pass through the origin. Their
slopes give a mean density of resonances of $\overline{\rho}=4.3\,$G$^{-1}$
for $^{164}$Dy, 2.7$\,\mathrm{G}^{-1}$ for $^{168}$Er at
$T=350\,$nK, 3.4$\,\mathrm{G}^{-1}$ for Er at $T=1.4$\,$\mu$K, and
25.6$\,\mathrm{G}^{-1}$ for $^{167}$Er. (c) Number variance
of the experimental data as a function of scaled $B$-field
interval $\overline{N}=\Delta B\overline{\rho}$. 
The experimental data for $^{164}$Dy
(blue line) and $^{168}$Er at $T=350$ nK (orange line) lies between
the variances for an uncorrelated Poisson distribution (dashed line)
and the correlated Wigner-Dyson distribution (dot-dashed line).
}
\label{fig:resonances} \end{figure*}

Chaotic behavior is manifest in a variety of complex systems ranging from
atomic to nuclear and solid-state physics.  In atomic physics chaos was
originally studied with Rydberg states of H and He in a magnetic field
\cite{Blumel1997}. Later on a variety of more complex atoms and ions
in highly-excited states have shown signatures of chaotic spectral
distributions \cite{Flambaum1998}.  The origin of chaos in these
systems was traced back to a strong mixing of many-electron excited
states by the Coulomb interaction \cite{Flambaum2014}. A chaotic level
distribution is also common in a variety of  solid state systems ranging
from those with strong many-body interactions to the motion of particles
in irregular potentials \cite{Altshuler,Zelevinsky}. Experiments in
nuclear physics \cite{Rahn1972,Wynchank1972} have also produced
substantial evidence for chaotic neutron resonance spectrum
fluctuations, which agree with predictions of random matrix theory
(RMT). Similar agreement was found from numerical simulations
based on nuclear shell models \cite{nshell,nshell2}.  Moreover,
Refs.~\cite{Weidenmuller2006,Weidenmuller2009} suggested that chaos is a
generic property of nuclei with multiple degrees of freedom (i.e. multiple
active shells), which become completely mixed.

This article describes a joint effort to understand ultracold scattering
and Fano-Feshbach spectra of strongly magnetic Er and Dy atoms.
In particular, we report on the measurement and statistical analysis of
Fano-Feshbach spectra for Dy and Er between $B=0\,$G to 70$\,$G at gas
temperatures $T$ below and around 1 $\mu$K.  Here 1$\,$G=0.1$\,$mT. We
observe that both elements have similar chaotic scattering. We present
a random-matrix-theory (RMT) inspired model to gain insight into
their statistical properties as well as theoretical evidence based on
coupled-channels calculations with a microscopic Hamiltonian that chaotic
scattering requires both strong molecular anisotropy and Zeeman mixing
to fully develop. Limitations of the RMT are also discussed.  Finally,
we present experimental data and a comparison to a resonant trimer model
to show that our increase in resonance density with temperature is a
consequence of the strong collision-energy dependence of transitions from
entrance $d$-wave channels of three free atoms to resonant trimer states.


\section{Experiment}
\subsection{Measurement}
The experimental study of Fano-Feshbach resonances in Er and Dy is based
on high-resolution trap-loss spectroscopy on  spin-polarized thermal
samples. Ultracold bosonic $^{164}\text{Dy}$
samples are created by direct loading from a narrow-line magneto-optical trap (MOT),
operating on the 626$\,$nm cycling transition, into a single-beam optical
dipole trap (ODT) \cite{DyMOT}. By moving the last focusing lens of
the ODT, the atoms are transported from the MOT chamber to the science
cell. The ODT is created with a 100$\,$W fiber laser at a wavelength
of 1070$\,$nm. We achieve a transport efficiency close to unity. This
fiber laser, however, causes atom loss due to its longitudinal multimode
structure \cite{IPGloss}. Therefore, we transfer the atoms into a second
single-beam ODT, created by a 55$\,$W solid-state laser at a wavelength
of 1064$\,$nm. Finally, forced evaporative cooling in a crossed ODT leads
to a sample of $10^5$ atoms in the energetically-lowest Zeeman sublevel
$m_j=-8$ at  $T=600\,$nK.

High-resolution trap-loss spectroscopy is performed
on a spin-polarized bosonic $^{168}\text{Er}$ sample at $T=1400\,$nK
and compares this spectrum with that obtained at a four times lower
temperature measured in previous work both for $^{168}\text{Er}$ as
well as for fermionic $^{167}\text{Er}$ \cite{ErstatFR}. The experimental
procedures for creating bosonic and fermionic samples are described in
Ref.~\cite{ErBEC} and \cite{ErFermion}, respectively. Bosons (fermions)
are prepared in the lowest Zeeman sublevel $m_j=-6$ ($m_f=-19/2$). Erbium
samples are trapped in a crossed ODT and contain about $10^5$ atoms.

\subsection{Feshbach spectroscopy}

Feshbach spectroscopy is performed in a similar manner for the two
species.  The magnetic field is ramped up over a few milliseconds to
a magnetic field value $B$, where the atoms are held in the ODT for
500$\,$ms for Dy, 400$\,$ms for $^{168}\text{Er}$, and 100$\,$ms for
$^{167}\text{Er}$. During this time  inelastic three-body recombination
causes atom loss from the ODT. At resonance, the recombination process
is enhanced because of the coupling between the atomic-threshold
state and a molecular state leading to a resonant increase of the atom
loss. We identify the field locations of maximum loss as the positions
of  Fano-Feshbach resonances \cite{ReviewFB}. The atom number is probed
by standard time-of-flight absorption imaging at low magnetic field. We
record atom-loss features for magnetic field values between 0$\,$G and
70$\,$G in steps of a few mG. Figure~\ref{fig:resonances}(a) shows the
normalized loss spectrum for the $^{164}\text{Dy}$ isotope, where  we identify
309 resonances. For $^{168}\text{Er}$  at $T=1.4\,\mu$K there are
238 resonances.  The Fano-Feshbach scan of fermionic $^{167}\text{Er}$
is carried out from 0$\,$G to 4.4$\,$G and yields 115 resonances.

The understanding of the richness of the scattering in Er and Dy requires
the development of sophisticated microscopic coupled-channels scattering
models. We defer such analysis until later in this paper and first analyze
our data following the statistical approach based on the random-matrix
theory (RMT) advocated by Ref.~\cite{ErstatFR}. In particular, we study
the correlations between resonance locations via the nearest-neighbor
spacing (NNS) distribution and setup a RMT-like model, which accounts
for the structure of our $B$-dependent microscopic Hamiltonian to get
intuition about these NNS. In our description of the coupled-channels
calculations limitations of such a RMT-like model are discussed.

\subsection{Statistical analyses}

Our statistical analysis starts with the construction of
the staircase function, which is a step-like function that
counts the number of resonances below magnetic-field value $B$
\cite{RMTnucl}. Figure~\ref{fig:resonances}(b) shows the staircase
function for Dy and Er. For both species the function is well-fit
by a linear curve forced to pass through the origin. Its slope $\bar
\rho$ corresponds to the density of resonances.  Deviations below and
above the fit occur for small and large $B$, respectively. The fitted
resonance densities are given in the caption.  Remarkably, the density
of resonances of $^{168}\text{Er}$ at $T=1.4\, \mu$K is 25$\,\%$ higher
than the one observed at 350$\,$nK. 
The discussion of the origin of this sensitivity is postponed until Section \ref{sec:Tdep}.
The density $\overline{\rho}$ for bosonic Dy is 50\% larger than for
bosonic Er. This is caused by the larger $\vec \jmath$ of Dy and,
thus, its larger number of allowed collision channels. The much larger
density $\overline{\rho}$ of 25.6$\,\mathrm{G}^{-1}$ for the fermionic
$^{167}\text{Er}$ is due to its additional hyperfine structure.

\begin{figure*}
\includegraphics[width=1.5\columnwidth]{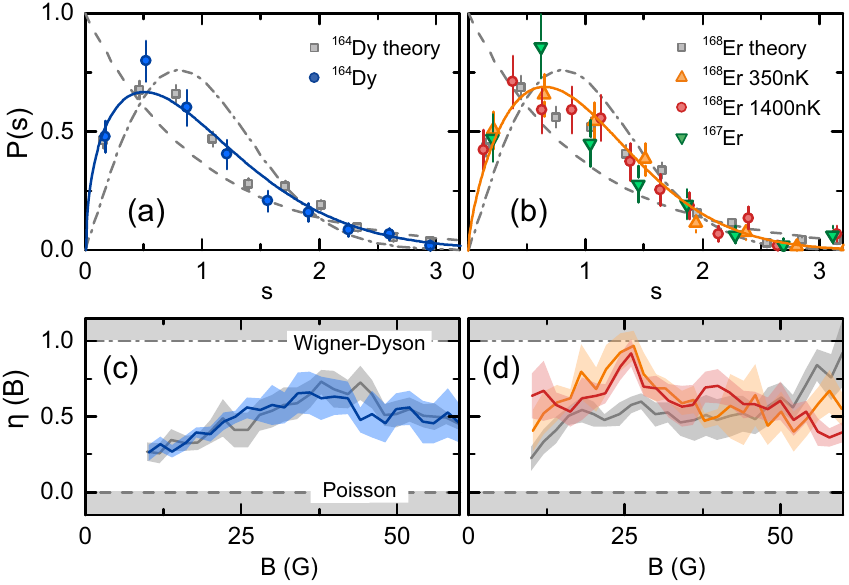}
\caption{(color online) (a) Nearest-neighbor spacing distribution,
$P(s)$, determined from all-observed Fano-Feshbach resonances for
$^{164}\text{Dy}$ (blue markers).  Dashed and dot-dashed curves
are the Poisson and Wigner-Dyson distribution, respectively.
The solid line is a Brody distribution with $\eta=0.45(7)$ fit to the
experimental data. (b) Distributions $P(s)$ for $^{168}\text{Er}$
at $T=350\,$nK and  $T=1.4$\,$\mu$K (orange and red markers, respectively),
and $^{167}$Er at $T=0.4\,T_F$ (green), where $T_F$ is the
Fermi temperature of the gas.  The solid line is a Brody distribution with
$\eta=0.68(9)$ fit to the $^{168}\text{Er}$ data at $T=350\,$nK.  Panels
(c) and (d) show the magnetic-field-resolved Brody parameter $\eta(B)$
as a function magnetic field for $^{164}$Dy and $^{168}$Er, respectively. The Brody parameters for the Poisson and Wigner-Dyson
distribution are 0 and 1, respectively. Gray markers and lines in all four panels
are results from our coupled-channels calculations. The $1\sigma$ error bars in (a)
and (b) correspond to Poisson counting errors while shaded bands in (c)
and (d) are $1\sigma$ statistical uncertainties of the fits
to the data.}
\label{fig:distribution} 
\end{figure*}

Fluctuations in the number of resonances within a magnetic field
interval $\Delta B$ is a second measure of the statistical properties
of the spacings between resonances.  Formally, it is defined as the
dimensionless number variance $\Sigma^2=\overline{N^2}-\overline{N}^2$,
where $\overline{N}=\sum_{i=0}^{M-1} N_i/M$, $\overline{N^2}=\sum_{i=0}^{M-1}
N^2_i/M$, and $N_i$ is the number of resonances in field interval [$i\Delta
B$, $(i+1)\Delta B$] with $i=0,\dots, M-1$, such that $M\Delta B= B_{\rm max}$
and $B_{\rm max}=70$ G for both species. Consequently, $\overline{N}\equiv\Delta B\overline{\rho}$. 
For shot noise or a Poissonian distribution
we expect $\Sigma^2=\overline{N}$.  Figure \ref{fig:resonances}(c)
compares $\Sigma^2$ for our Dy and Er data as a function of $\overline
N$.  The fluctuations for both species monotonically increase with
$\Delta B$ but are  substantially less than the shot noise limit. 
While this behavior was previously demonstrated for
Er \cite{ErstatFR}, the present results provide the first evidence of
correlation in Dy and indicate similarity between the species.

These correlations between resonance locations are further studied
using the nearest-neighbor spacings distribution $P(s)$ where $s=\delta
B\,\overline{\rho}$ and $\delta B$ is the field spacing between two adjacent
resonances in the spectrum. Figures \ref{fig:distribution}(a)
and (b) show the computed NNS distribution of our experimental
data, derived from the number of NNS, $S_i$, with spacings $s$
between $i\delta s$ and $(i+1)\delta s $, where $i=0,1,\dots$ and $\delta s
\approx0.3$. The NNS distributions have clear deviations from both
the Poisson $P_\mathrm{P}(s)=\exp(-s)$ and Wigner-Dyson $P_{\mathrm{WD}}(s)=(\pi/2)s\exp[-(\pi/4)s^2]$
distribution, two well-known distributions within RMT \cite{ErstatFR}.
A Poisson distribution corresponds to a random distribution of
resonance locations, while a Wigner-Dyson distribution corresponds
to a situation, where neighboring resonances ``avoid'' each other
and $P_{\mathrm{WD}}(s)\propto s$ for $s\to 0$.  Deviations are
also seen in Fig.~\ref{fig:resonances}(c), where for both atomic
species the variance $\Sigma^2$ does not agree with the corresponding
predictions for these distributions.  The experimental NNS distributions
in Figs.~\ref{fig:distribution}(a) and (b) have also been fit to the
Brody distribution $P_\mathrm{B}(s,\eta)=b (1+\eta)s^\eta \exp[-bs^{\eta+1}]$, an empirical function that
interpolates between $P_\mathrm{P}(s)$ and $P_{\mathrm{WD}}(s)$ for
$\eta=0$ and 1, respectively, and $b$ is a normalization constant \cite{RMTstat}. The values for $\eta$
reported in the caption, indicate intermediate or mixed behavior of the data.

We present the magnetic-field resolved Brody parameter $\eta(B)$  in
Figs.~\ref{fig:distribution}(c) and (d) obtained from a fit to the
NNS distribution of resonances located in moving intervals $[B-\Delta
B/2,B+\Delta B/2]$ with $\Delta B=20$ G. It has a non-negligible
$1\sigma$ uncertainty equally limited by the quality of the fit and
the number of Feshbach resonances in an interval or bin. The latter
uncertainty is reflected in the bin-to-bin variation of $\eta(B)$.
For Dy we observe that $\eta$ increase linearly with field for small
$B$, which saturates at a value of  $\approx 0.5$ for $B>30$
G. For Er the Brody parameter fluctuates around 0.5. Interestingly, the
Er data at our two temperatures have a similar behavior, indicating that
the larger density of resonances at higher $T$ does not impact
the degree of correlation between their spacings.

\begin{figure}[b]
  \includegraphics[width=1.1\columnwidth]{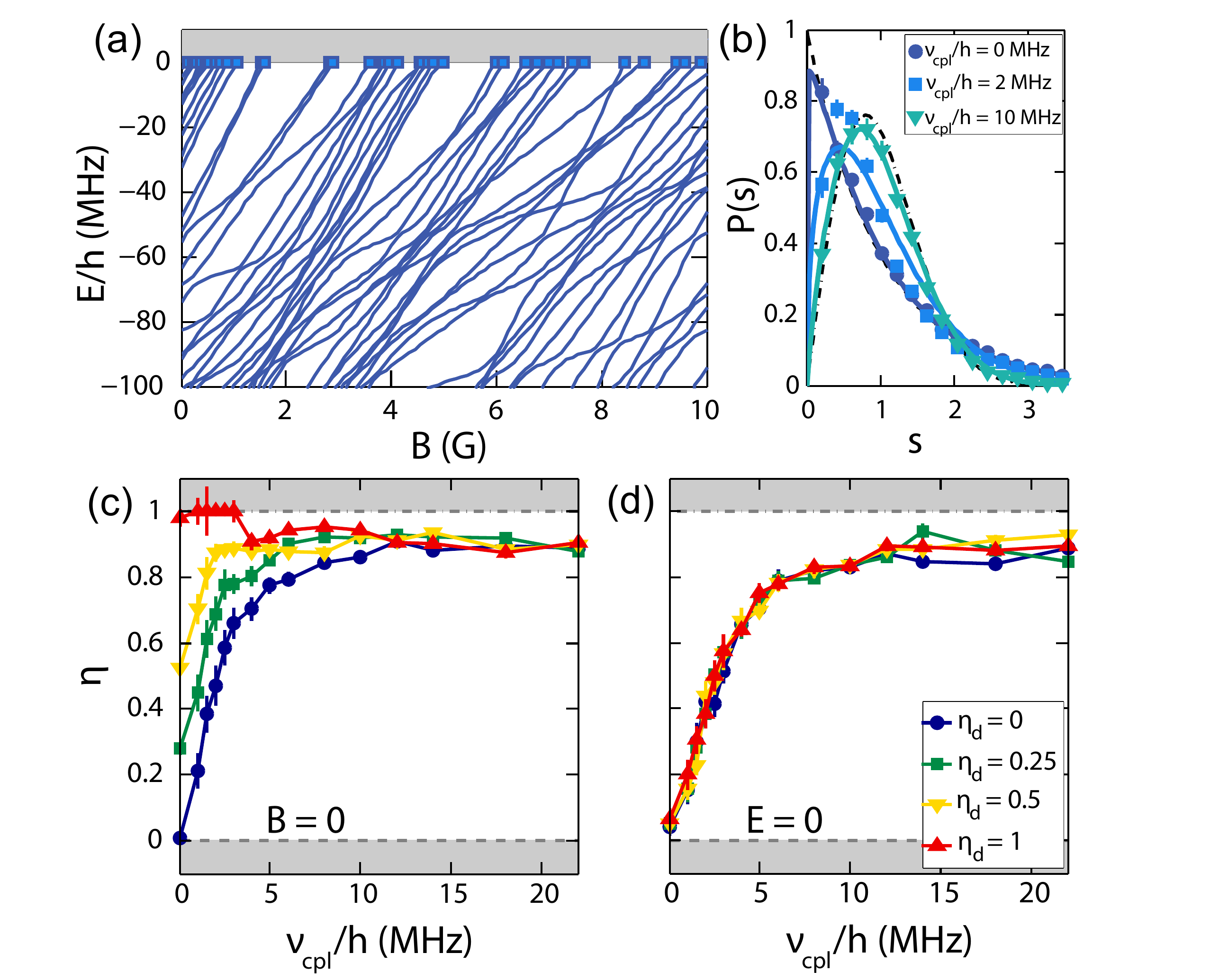}
  \caption{(color online) 
   Molecular spectrum and NNS distributions of
  Fano-Feshbach resonances of $^{168}$Er calculated from our RMT model.
  (a) Example of a spectrum of molecular binding energies as a function
  of $B$ for $\nu_{\rm cpl} /h = 2$ MHz, $\eta_{\rm d}=0$. Here, $h$ is Planck's constant.
  Squares at $E/h=0$ MHz indicate crossings of
  molecular levels with the threshold of two $m_j=-6$ atoms
  and correspond to the position of Feshbach resonances. For the sake
  of visibility we only show the spectrum between $B=0\,$G to 10\,G.
  (b) NNS distributions of simulated Feshbach resonances for $\eta_{\rm
  d}=0$ and $\nu_{\rm cpl}/h=0$ MHz (circles), $ 2\,\mathrm{MHz}$ (squares),
  and $10\,\mathrm{MHz}$ (triangles). The dashed and dashed-dotted lines
  are Poisson and Wigner-Dyson distributions, respectively. Solid lines
  are best-fit Brody distributions with $\eta = 0.03(1)$, $\eta=0.41(5)$
  and $\eta=0.82(1)$ for $\nu_{\rm cpl}/h=0$ MHz, $ 2\,\mathrm{MHz}$,
  and $10\,\mathrm{MHz}$, respectively.  (c) Fitted Brody parameters of
  the NN energy-spacing distribution of the eigenvalues of $H_{\rm RMT}$
  at $B=0$ G as a function of $\nu_{\rm cpl}$. Circles, squares, down
  triangles, and up triangles correspond to $\eta_{\rm d}=0$, $0.25$,
  $0.5$, and $1$, respectively.  (d) Fitted Brody parameters of the
  NNS distribution of the Feshbach resonances as a function of $\nu_{\rm cpl}$
  and for four $\eta_{\rm d}$ using the same marker code as in (c).
  In (b), (c), and (d) distributions are obtained by averaging over
  15 realizations of $H_{\rm RMT}$, each of dimension $500\times500$
  and using Feshbach resonances computed up to $B=85\,$G.
	}
  \label{fig:couplingmodel}
\end{figure}

\section{ RMT ensemble model}

Random matrix theory is based on the powerful notion that the statistics
of eigenvalues and eigenfunctions of a complex system can be studied
by replacing the microscopic Hamiltonian by an ensemble of random
Hamiltonians.  In this spirit, we construct a RMT-inspired model
for weakly-bound molecular dimer states to test the distribution of
Fano-Feshbach resonances.

Our RMT model is based on the statistics of eigenvalues of the $N\times
N$ real, symmetric matrix $H_{\rm RMT}=H_0+H_{\rm Z}$, where matrices
$H_0$ and $H_{\rm Z}$ represent the $B=0$ Hamiltonian and the Zeeman
interaction of the two atoms, respectively.  Without loss of generality
we can assume that $H_{\rm Z}$ is a diagonal matrix with matrix elements
given by $mg\mu_B B$, where $m$ is an integer between $-2j$ and $2j$,
corresponding to the sum of the projection quantum numbers of the atomic
angular momenta, $g$ is the atomic Land\'e factor, and $\mu_B$ is the
Bohr magneton.  The Zeeman interaction does not depend on the rotational
state of the molecule and, thus, entries in $H_{\rm Z}$ correspond to
states with a definite value for $\ell$ and its projection. $H_0$ is then
the $B=0$ Hamiltonian expressed in this basis. It is also convenient to
define $H_0=H_{\rm d}+H_{\rm cpl}$, where diagonal matrix $H_{\rm d}$
contains the diagonal matrix elements of $H_0$ and $H_{\rm cpl}$ is the
matrix of all its off-diagonal elements. The eigenvalues of $H_{\rm d}$
can then be interpreted as the energies of ro-vibrational levels of the
isotropic contribution of the molecular BO potentials, while $H_{\rm cpl}$
describes mixing due to the anisotropic contributions of these potentials.

We generate members of our ensemble of $H_{\rm RMT}$ by choosing
random matrix elements for $H_{\rm Z}$, $H_{\rm d}$, and $H_{\rm cpl}$
based on specific distributions.  The values of $m$ in $H_{\rm Z}$
are uniformly distributed integers between $-2j$ and $2j$.  The matrix
elements of $H_{\rm d}$ are chosen according to a Brody distribution
with variable Brody parameter $\eta_{\rm d}\in [0,1]$ and with a mean
energy spacing between bound states $\epsilon_{\rm d}$.
Finally, matrix elements of $H_{\rm cpl}$ are chosen as
Gaussian-distributed real numbers with zero mean and standard deviation
$\nu_{\rm cpl}$, thereby on average coupling all diagonal elements
equally.  Notice that this construction deviates from that for a true
Gaussian orthogonal ensemble, where all matrix elements of a symmetric
Hamiltonian are Gaussian distributed \cite{RMTens}.

We apply the RMT model to the case of $^{168}$Er. The relevant species-specific 
quantities are $j=6$, $g=1.16$, and $\epsilon_{\rm d}$ is chosen to  roughly reproduce the
observed density of Fano-Feshbach resonances of $^{168}$Er
and is set to $\epsilon_{\rm d}/h = 6.4$\, MHz, where h is Planck's constant.
Figure \ref{fig:couplingmodel}(a) shows an example of a molecular
spectrum, the eigenvalues of $H_{\rm RMT}$ obtained with
our RMT model as a function of $B$ with $\eta_{\rm d}=0$ and $\nu_{\rm
cpl}/h=2$ MHz.  We observe that as $B$ increases weakly-bound
molecular states avoid each other multiple times before reaching the
two-atom threshold creating a Feshbach resonance. When we turn off
$H_{\rm cpl}$ the levels cross.  Similar $B$-field dependencies of the
eigenvalues occur for $\eta_{\rm d}>0$.

We investigate the effect of the parameters $\nu_{\rm cpl}$ and $\eta_{\rm
d}$ on the NNS distribution of the Fano-Feshbach resonances as well as
that of the $B=0$ molecular levels.  Figure \ref{fig:couplingmodel}(b)
shows the NNS distribution of Feshbach resonances, obtained by averaging
over 15 realizations of $H_{\rm RMT}$, for four values of $\nu_{\rm cpl}$
and $\eta_{\rm d}=0$.  For negligible $\nu_{\rm cpl}$ the distribution
follows $P_\mathrm{P}(s)$ and approaches $P_\mathrm{WD}(s)$ when the
anisotropic coupling strength $\nu_{\rm cpl}$ is large compared to
$\epsilon_{\rm d}$. In fact, we find that a larger $\epsilon_{\rm d}$
requires a larger $\nu_{\rm cpl}$ to develop correlations.

Figures \ref{fig:couplingmodel}(c) and (d) show Brody parameters fit
to NNS distributions as functions of $\nu_{\rm cpl}$ and $\eta_{\rm
d}$. Panel (c) shows $\eta$ for the $B=0$ molecular binding energies. For
$\nu_{\rm cpl}=0$ the Brody parameter is simply $\eta_{\rm d}$ as
expected from the distribution of the diagonal $H_{\rm d}$, while for
larger interaction anisotropy $\nu_{\rm cpl}$ the parameter $\eta\approx
0.9$, close to a Wigner-Dyson distribution, independent of $\eta_{\rm d}$.
Panel (d) shows $\eta$ extracted from the RMT Feshbach resonance locations
as a function of $\nu_{\rm cpl}$. It suggests that the correlation in the
NNS of the resonances is caused by $\nu_{\rm cpl}$, whereas it appears
fairly independent on $\eta_{\rm d}$. More precisely, the Brody parameter
fitted to these distributions rapidly increases from $\eta \approx 0$
to  $\eta \approx 0.8$ for $\nu_{\rm cpl} \lesssim \epsilon_{\rm d}$
and tends to one for larger $\nu_{\rm cpl}$.  
We conclude from the RMT model that the correlations between the locations of the Fano-Feshbach
resonance are essentially due to the avoided crossings between
weakly-bound molecular states at finite $B$ and only weakly dependent on the energy distribution at $B=0$. In fact, these correlations increase for increasing $\nu_{\rm cpl}$.


\section{Microscopic coupled-channels model}\label{sec:micro}

\subsection{Realistic setup}\label{sec:realis}

A quantitative understanding of the origin of the chaotic resonance
distribution requires coupled-channels and bound-state calculations
with physically realistic angular-momentum couplings and interaction
potentials.  We do so here based on the time-reversal symmetric
Hamiltonian for the relative motion of  Dy and Er described
in Refs.~\cite{Petrov2012,ErstatFR}. It contains the Zeeman
Hamiltonian, the molecular vibration and rotation, and the molecular
interactions with isotropic (orientation-independent) and anisotropic
(orientation-dependent) contributions,  $\hat V_{\rm i}(R)$ and $\hat
V_{\rm a}(\vec R)$, respectively, where $\vec R$ describes the separation
$R$ and orientation of the atom pair $\hat R$. The potential has eight  tensor
operators coupling the two  atomic and relative orbital angular momenta,
$\vec \jmath_1$, $\vec \jmath_2$, and $\vec\ell$. For $B=0$  the total
angular momentum $\vec J= \vec \jmath_1+\vec \jmath_2+\vec\ell$ is
conserved.  For $B>0$ G only the projection $M$ of $\vec J$ along $\vec B$
is conserved.  The zero-of-energy of the Hamiltonian is the energy of an
atom pair  in the absolute lowest Zeeman sublevel $m_{j\alpha}=-j_\alpha$.

The potentials $\hat V_{\rm i}(R)$ and $\hat V_{\rm a}(\vec R)$
contain short-ranged exchange, medium-ranged van der Waals as well
as long-range magnetic dipole-dipole interactions.  We use the
isotropic van der Waals coefficient $C_6=1723\,E_ha_0^6$ and
anisotropic coefficients spread over $\Delta C_6=174\,E_ha_0^6$ for Er
\cite{ErstatFR}. For Dy we have improved the value of van der Waals
coefficients of Ref.~\cite{Petrov2012} by including additional
experimental and theoretical transition frequencies and oscillator
strengths \cite{DyTrans1,DyTrans2,DyTrans3,DyTrans4} and now use
$C_6=2003\,E_ha_0^6$ and spread $\Delta C_6=188\,E_ha_0^6$.  In particular, the
anisotropic spread for Dy has significantly increased. Here, $E_h=4.360\times
10^{-18}$ J is the Hartree energy and $a_0=0.05297$ nm is the Bohr
radius. 

\subsection{Bound-state calculations}

In Ref.~\cite{ErstatFR} we performed initial coupled-channels
calculations of the scattering between ultracold Er atoms and
predicted that tens of partial waves $\ell$ should have been included
as the strength of the anisotropic contribution is large. We,
however, were unable to reach numerical convergence with respect
to the number of coupled equations. 

Here, we circumvent this limitation by performing multichannel bound-state
calculations, in which we use $B=0$ eigenstates as a basis for those at
$B>0$ G. For $B=0$, where $J$ is a good quantum number, at most 49 and 81
Bose-symmetrized and parity-conserving channels are coupled for Er and Dy,
respectively.  The $B=0$ coupled Schr\"odinger equations are discretized on
the interval $R\in[0,R_{\rm max}]$ assuming zero boundary conditions and
solved as a matrix eigenvalue problem
\cite{Tiesinga1996,Tiesinga1998,Kokoouline2000,Colbert1992}.  For each $J$
only eigenstates with energies between $[E_0,E_1]$ surrounding the zero of
energy are computed and stored.  The bound states for $B>0$ G are solutions
of the matrix eigenvalue problem that includes all computed $B=0$  solutions
with $|M|\le J\le J_{\rm max}$ and their coupling due to the Zeeman
interaction.  Selection rules of the Zeeman interaction ensure that there
only exists direct coupling between $J$ and $J'$ zero-field eigenstates with
$J-J'=0,\pm 1$.  For both species
$R_{\rm max}=1000\,a_0$, $E_0/h=-3 $ GHz and $E_1/h=0.9$ GHz ensuring
that Feshbach resonance locations below 70 G are converged.

In this section on the microscopic calculations we focus on analysing
the spectra at our coldest temperatures, where the initial collision
channel has $s$-wave ($\ell=0$) character. Hence, we only need to consider
even-$\ell$ channels with total projection quantum number $M=-12$
and $-16$ for $^{168}$Er and $^{164}$Dy, respectively and inclusion
of zero-field solutions up to $J_{\rm max}=36$ for Dy and $39$ for
Er is sufficient to reproduce the experimental resonance densities.
In Sec.~\ref{sec:Tdep} we will discuss higher temperature collisions 
between Er atoms, where $d$-wave ($\ell=2$) entrance channels must be 
considered and, hence, spectra at other $M$ values (i.e. $M$ between 
$-14$ and $-10$ for $^{168}$Er) contribute.

\begin{figure*}
\includegraphics[scale=0.25,trim=0 0 0 0,clip]{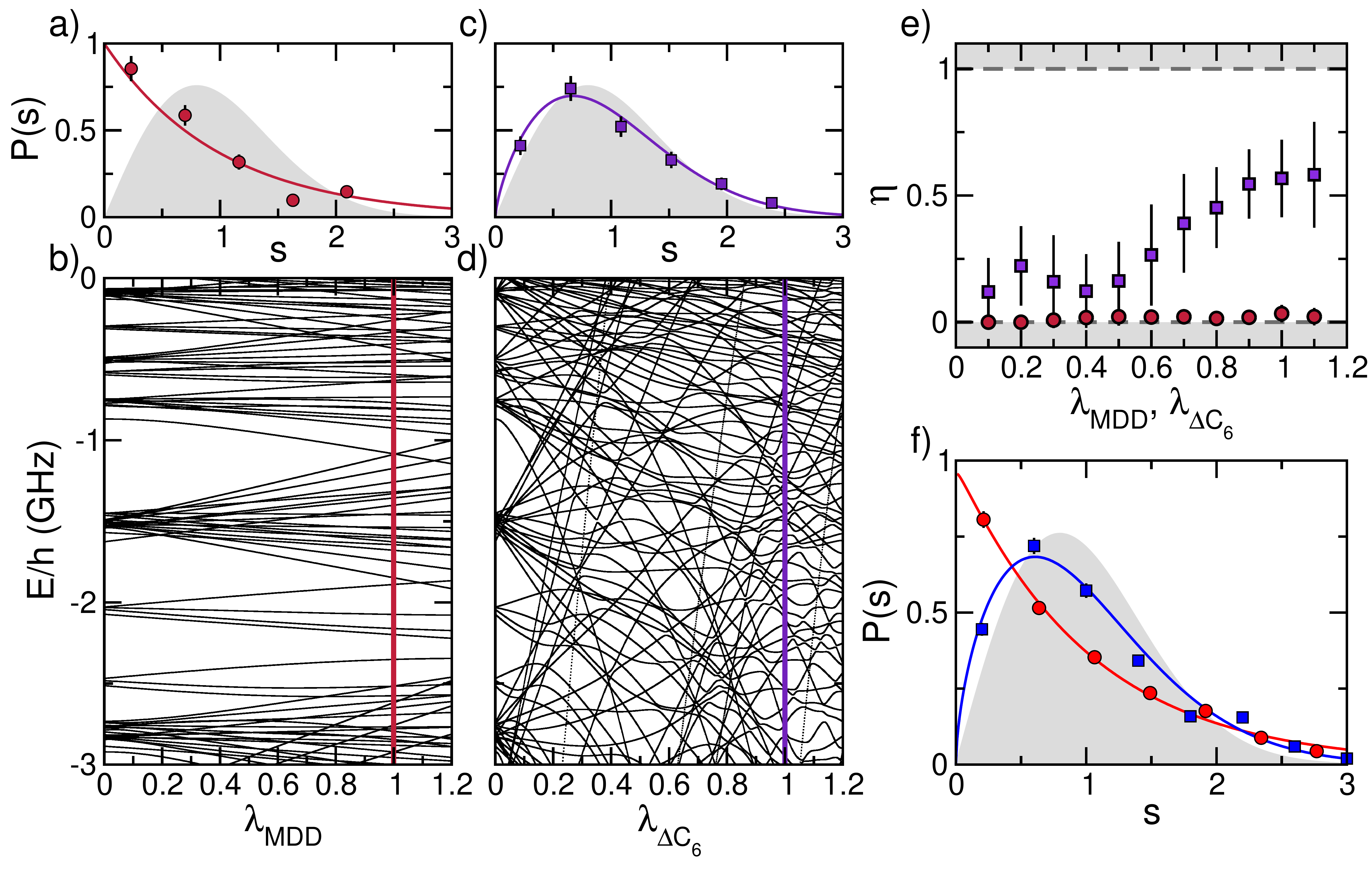}
\caption{ (Color online) Interaction anisotropy induced chaos of
$B=0$  near-threshold bound states. 
(b) Weakly-bound
$J=16$ bound-state energies  of $^{164}$Dy$_2$  as a function of
the anisotropy scale $\lambda_{\rm MDD}$ with $\lambda_{\Delta C_6}=0$.  
(a) NNS  distribution (red circles) for the $J=16$ bound state data in panel (b) at $\lambda_{\rm MDD}=1$ and 
$\lambda_{\Delta C_6}=0$. The solid red line is a Brody distribution fit to the data and agrees well with a Poisson distribution.
(d) Weakly-bound
$J=16$ bound-state energies  of $^{164}$Dy$_2$  as a function of
the anisotropy scale $\lambda_{\Delta C_6}$ with $\lambda_{\rm MDD}=0$.  
(c) NNS  distribution (purple squares) for the $J=16$ bound state data in panel (d) at $\lambda_{\rm MDD}=0$ and 
$\lambda_{\Delta C_6}=1$. The solid purple line is a Brody distribution fit to the data and is close to a Wigner-Dyson distribution.
(e) Moving average of the Brody parameter $\eta$ as a function of $\lambda_{\Delta C_6}$ (purple squares) or 
$\lambda_{\rm MDD}$ (red circles) with bins $\Delta\lambda=0.2$ obtained by fitting  the NNS  distribution  for
the $J=16$ bound state data in panel (b) and (d) to Brody distributions, respectively.  The horizontal
lines at $\eta=0$ and 1 correspond to the Brody parameter for a Poisson  and Wigner-Dyson distribution, respectively.  
The $1\sigma$ error bars  combine  statistical and fitting uncertainties.  
(f) The individual-$J$ (blue squares)
and combined-$J$ (red circles) NNS distributions $P(s)$ at $\lambda_{\rm MDD}=\lambda_{\Delta C_6}=1$
as a function of the normalized energy spacing $s$.  The distributions are derived from $B=0$ bound-state data for $J=16,\dots,25$.
The grey shaded areas in panels (a), (b), and (f) indicate the Wigner-Dyson distribution.
} \label{fig:DyAnisoPlotsA}
\end{figure*}

\subsection{Interaction anisotropies}

We first look into the role of interaction anisotropies on the level
distribution of the most-weakly bound molecular energy levels at zero magnetic
field. There are two dominant components to the anisotropy, the dispersion  $V_{\Delta C_6}(\vec R)$ and magnetic
dipole-dipole $V_{\rm MDD}(\vec R) $ contribution. To distinguish the contributions of these two terms, we define:
\begin{equation}
   \hat V_{\rm a}(\vec R)=\lambda_{\Delta C_6}V_{\Delta C_6}(\vec R)+\lambda_{\rm MDD} V_{\rm MDD}(\vec R) 
\end{equation}
with variable strength $\lambda_{\Delta C_6}$ and $\lambda_{\rm MDD}$.
We systematically increase the strengths  $\lambda_{\Delta C_6}$ and $\lambda_{\rm MDD}$
from zero, where we recover the full physical strength for $\lambda_{\rm MDD}=\lambda_{\Delta C_6}=1$. 

For completeness we note that the dominant tensor operator for the anisotropic dispersion contribution is
\[
V_{\Delta C_6}(\vec R) = \frac{c_a}{R^6} \sum_{i=1,2} \frac{1}{\sqrt{6}}\left\{3(\hat R \cdot \vec \jmath_i)(\hat R \cdot \vec \jmath_i)-\vec \jmath_i \cdot \vec \jmath_i   \right\}+ \cdots
\]
with strength $c_a<0$  found with the methodology discussed in the subsection \ref{sec:realis}. Weaker contributions indicated
by dots are included in our calculations.
Moreover,
\[
V_{ \rm MDD}(\vec R) = -\frac{\mu_0}{4\pi} \frac{(g\mu_B)^2}{R^3} \left\{3(\hat R\cdot \vec \jmath_1)(\hat R\cdot \vec \jmath_2) -\vec \jmath_1\cdot \vec \jmath_2 \right\} \,,
\]
where $\mu_0$ is the magnetic constant.

Figure \ref{fig:DyAnisoPlotsA}(b) and (d) show the most-weakly-bound $B=0$, $J=16$ levels of  $^{164}$Dy$_2$ as a function of anisotropy strength for purely dipolar ($\lambda_{\Delta C_6}=0$, varying $\lambda_{\rm MDD}$) and dispersive ($\lambda_{\rm MDD}=0$, varying $\lambda_{\Delta C_6}$) anisotropic interaction, respectively.
For $\lambda_{\rm MDD}=\lambda_{\Delta C_6}=0$ the binding energies
are  regularly structured with many near degeneracies. In fact, the
corresponding states are  ro-vibrational levels of the isotropic
centrifugal potentials $\hat V_{\rm i}(R)$ and labeled by $\ell$.
In our $3$ GHz energy window an $s$-wave channel has at most three
bound states, while even $\ell>0$ channels with their centrifugal
barriers have fewer \cite{MQDT,ErstatFR}.  For small  $\lambda_{\Delta C_6}$ and $\lambda_{\rm MDD}$ the
degeneracy is lifted and levels shift linearly.
The linear dependence for increasing strength of the dipole-dipole is approximately valid
up to the physical value of $\lambda_{\rm MDD}=1$.
Hence, the dipole-dipole interaction does not lead to our chaotic level distributions.
In fact, Fig.~\ref{fig:DyAnisoPlotsA}(a) shows that at $\lambda_{\rm MDD}=1$ and $\lambda_{\Delta C_6}=0$
the NND distribution is Poissonian.

\begin{figure*}
\includegraphics[scale=0.38]{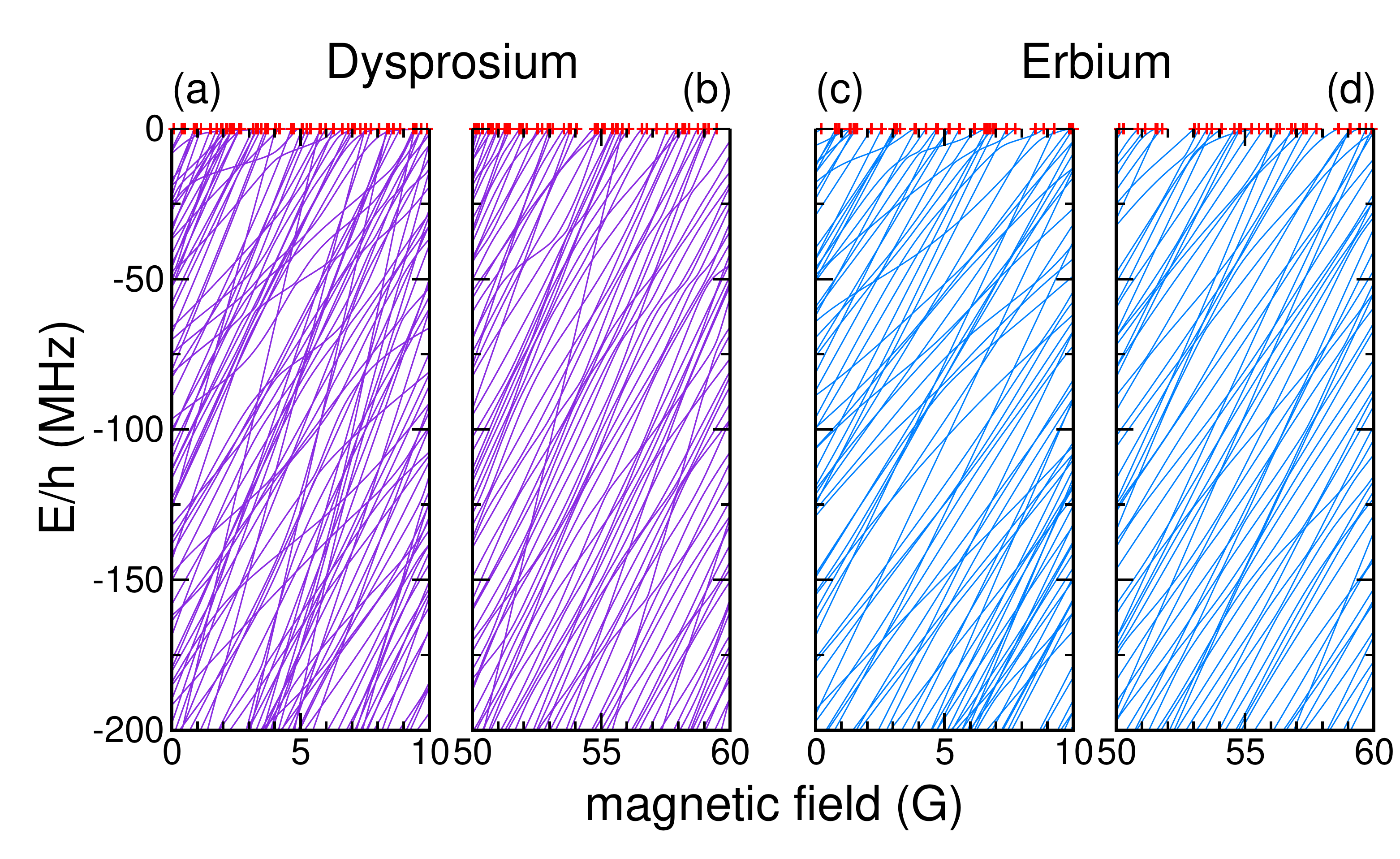}
\caption{(Color online) Theoretical $^{164}$Dy$_2$ (panels (a) and (b)) 
and $^{168}$Er$_2$ (panels (c) and (d)) near-threshold bound states as
a function of magnetic field.  Calculations have been performed
with the full nominal anisotropy.  Panels (a) and (c) show the
near-threshold region between $B = 0$ G to 10 G while panels (b) and (d)
show the region between $B = 50$ G to 60 G. Red crosses indicate the
location of Feshbach resonances. } 
\label{fig:ErDy_lines_HighLow}
\end{figure*}

On the other hand, for a relatively small anisotropic dispersion strength $\lambda_{\Delta C_6}\approx 0.1$  levels start to 
avoid each other. Starting from  $\lambda_{\Delta C_6}\approx  0.5$ most avoided crossings are
noticeable on the $3$ GHz scale of the figure.  At the nominal
$\lambda_{\Delta C_6}=1$, where there are 56 levels with  $-3$ GHz $<E/h< 0$
GHz, a significant fraction of the levels have undergone multiple
avoided crossings and  can not be described by a single dominant partial wave.  
The level spacing is chaotic as confirmed by the NND distribution for $\lambda_{\Delta C_6}=1$ and $\lambda_{\rm MDD}=0$ in Fig.~\ref{fig:DyAnisoPlotsA}(c).
We have  computed the weakly-bound $J=16$ levels for $\lambda_{\rm MDD}=\lambda_{\Delta C_6}=1$.
Visually the level distribution is much the same as the one shown in Fig.~\ref{fig:DyAnisoPlotsA}(d).
Similar results have been obtained for $^{168}$Er$_2$.

\begin{figure*}[t]
\includegraphics[scale=0.45]{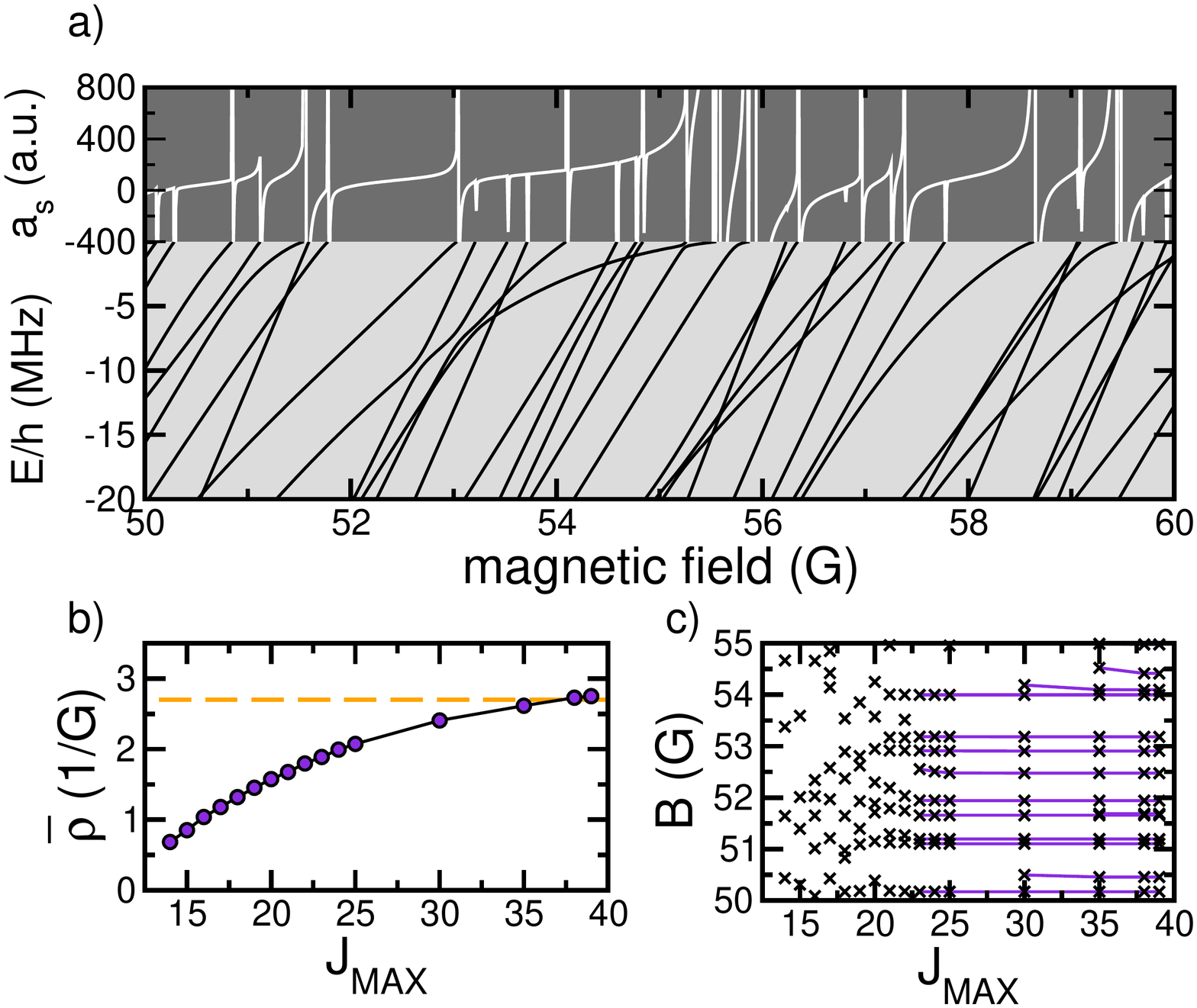}
\caption{(Color online) Theoretical near-threshold bound states and Feshbach resonances in a magnetic field.
(a) Near-threshold bound states for $M=-12$ $^{168}$Er$_2$  for  fields between $B=50$ G and 60 G 
(bottom half of image). 
The top half of the image shows the effective scattering length,  defined in the text, as a function of $B$.
It is infinite at a resonance location where a bound state has zero energy.
Calculations  use the physical interaction anisotropies and channel states with $J\le J_{\rm max}=39$.
(b) Theoretical Feshbach resonance density $\bar \rho$  as a function of $J_{\rm max}$ (purple circles)
computed from resonance locations between 0 G and 70 G.
The dashed horizontal  line  indicates the experimental density for $^{168}$Er$_2$. 
(c) Convergence study of resonance locations (crosses) between 50 G and 55 G as a function of $J_{\rm max}$.
Purple lines connect resonances when their location has converged.
} \label{fig:Er_convergence}
\end{figure*}

Figure~\ref{fig:DyAnisoPlotsA}(e) quantifies the intuition gained from panels (a)-(d) by showing the Brody parameter $\eta$ of
the $B=0$ $J=16$ $^{164}$Dy$_2$ levels as a function of $\lambda_{\Delta C_6}$ or $\lambda_{\rm MDD}$.
The Brody parameter is obtained by fitting a Brody distribution to the NNS distribution of the bound state data in 
Fig.~\ref{fig:DyAnisoPlotsA}(b) and (d).  
For increasing dipole-dipole strength $\lambda_{\rm MDD}$ and no anisotropic dispersion ($\lambda_{\Delta C_6}=0$) 
the parameter is always zero indicating the prevalence of small level spacings.
On the other hand, in the absence of the dipole-dipole interaction, increasing  $\lambda_{\Delta C_6}$ leads to an increasing $\eta$. It evolves from $\eta=0.2$ for $\lambda_{\Delta C_6}\lesssim 0.5$ to  $\eta=0.7$ for $\lambda_{\Delta C_6} =1$, indicating a depopulation of small energy spacings. Note that our systems does not reach a Wigner-Dyson distribution, which corresponds to $\eta=1$.

We compare in Fig.~\ref{fig:DyAnisoPlotsA}(f) two NNS distributions of $B=0$ weakly-bound states of  
$^{164}$Dy$_2$ obtained for the full anisotropic interaction ($\lambda_{\rm MDD}=\lambda_{\Delta C_6}=1$).
Both distributions are based on  $|E/h|< 3$ GHz bound states computed for $J=16$ up to $25$. 
The first so-called individual-$J$  distribution is constructed 
by averaging the NNS distribution of levels for  individual $J$s assuming that individual distributions are the same.  
The second, combined-$J$ NNS distribution, is calculated from a sorted list of all $J=16,\dots,25$ levels.
Data for $J>25$  are not included as the number of bound states is too small for a
reliable determination of the NNS distribution.  

The individual-$J$ NNS distribution is non-Poissonian as
levels with the same $J$ repel each other.  The combined-$J$ distribution, however,
 follows a Poisson distribution indicating that  energies of
bound states with different $J$ are uncorrelated. In other words,
even though the Hamiltonian, i.e. the set of coupling operators
between $\vec \jmath_1$, $\vec \jmath_2$
and $\vec \ell$, is the same,  differences in the matrix elements
and thus coupling strengths between channels lead to uncorrelated
eigen energies.

\subsection{Atom scattering in a magnetic field} 

The study of the $B=0$ G multi-channel bound states has shown that
interaction anisotropies mix channels with the same $J$, while states with
different $J$ are uncorrelated. The Zeeman interaction  mixes these
molecular levels  and leads to the Fano-Feshbach spectrum.  Figures
\ref{fig:ErDy_lines_HighLow}(a) and (b) show example $^{164}$Dy$_2$ $M=-16$ bound-
state spectra as a function of $B$ on two binding-energy and field regions.
Similarly, Figs.~\ref{fig:ErDy_lines_HighLow}(c) and (d) show $M =-12$ $^{168}$Er$_2$
bound states. In all cases the full nominal anisotropy ($\lambda_{\Delta C_6}=1$ and $\lambda_{\rm MDD}=1$) is used. For Dy and
Er channels with $J$ up to 36 and 39 are included, respectively. The figure
shows that the Dy level density is higher than that for Er. This simply
follows from the larger atomic angular momentum of Dy, leading to a larger
number of channels with the same $J-|M|$. We also observe that for both
species the level structure in the 0 G to 10 G, small field region is
qualitatively different from that in the larger field region. For small B
the avoided crossings are substantially narrower than for larger B.
Moreover, at small field the levels cluster, while at larger field they are
more uniformly distributed. These changes are a consequence of the linearly-increasing Zeeman coupling between vibrational levels with different $J$s as a
function of B.

Figure \ref{fig:Er_convergence}(a) shows effective length $a_s(B)$
as a function of $B$. It diverges at every resonance location and is closely related to 
the scattering length of a zero-energy collision. Our calculations can not be directly used to define the 
scattering length as we use a hard-wall potential for $R\ge R_{\rm max}$.  This
wall leads to a discrete set of  states with positive energy and using the lowest of these $E_s(B)$ we can 
define the effective  length $a_s(B)$ shown in the figure by solving for $E_s(B)=\hbar^2\pi^2/[2\mu_r (R_{\rm max}-a_s(B))^2]$
with $\mu_r=m/2$ and atomic mass $m$ \cite{Burnett2002}.
 
It is of interest to briefly discuss the convergence properties of our calculations.
The data in Figs.~\ref{fig:Er_convergence}(a) and \ref{fig:ErDy_lines_HighLow}(c) and (d) are based on computations with channels 
with $J$ up to $J_{\rm max}=39$.  Figure \ref{fig:Er_convergence}(b) shows the $^{168}$Er$_2$ Feshbach resonance 
density $\bar\rho$ as a function of $J_{\rm max}$. The resonance density increase linearly from $\approx 0.5$ 1/G at 
$J_{\rm max}=12$ but then is seen to ``saturate'' for larger $J_{\rm max}$. At $J_{\rm max}=39$ the experimental density 
is reproduced. In addition, Fig.~\ref{fig:Er_convergence}(c) shows  the field location of  resonances between 50 G and 
55 G as a function of $J_{\rm max}$. The resonance locations change significantly for $J_{\rm max}<22$, but
then rapidly converge. This implies strong mixing among bound states with those $J$. On the other hand, the location of 
resonances that appear for $J\ge22$ are almost immediately converged indicating weak mixing to smaller $J$ states.

\subsection{Comparison of experiment and coupled-channels model}

In Figs.~\ref{fig:distribution}(a) and (b) we show the NNS distribution
of converged Feshbach-resonance locations based on our multi-channel 
data between $B = 0$ G and 70 G for $^{164}$Dy$_2$ with $J_{\rm max}=36$ and 
$^{168}$Er$_2$ with $J_{\rm max}=39$, respectively. For both species the distribution clearly deviates from a Poisson distribution, consistent with the
experimental distributions that are also shown.  The fitted experimental
and coupled-channel Brody parameters agree within their error bars.

The anisotropy parameters $\lambda_{\Delta C_6}$ and $\lambda_{\rm MDD}$ in the coupled-channels calculations
and the parameter $\nu_{\rm cpl}$ in the RMT play analogous
roles in the Hamiltonian and in the emergence of chaotic level distributions, even though no
explicit quantitative connection exists.  This role is most manifest
in the Brody parameters of the $B=0$ G bound states and that of the
Feshbach resonance spectra for the two models.  For $^{168}$Er the
corresponding Brody parameters from the coupled-channels calculations are
$\approx\, 0.01$ and 0.68 at the physical $\lambda_{\rm MDD}=\lambda_{\Delta C_6}=1$, respectively.
Within the RMT model the small $\eta$ value for the $B=0$ G level
distribution requires weak coupling $\nu_{\rm cpl}\ll \epsilon_{\rm d}$
and $\eta_{\rm d}\approx0$.  In contrast the Brody parameter for the
Feshbach resonance spectrum requires $\nu_{\rm cpl}\approx \epsilon_{\rm
d}$  and points at limitations of the current RMT model.
Similar conclusions hold for bosonic Dy.  Future advanced RMT models
might circumvent these limitations by incorporating overlapping, uncoupled
chaotic series as is found from our $B=0$ G coupled-channels calculations.

We plot the $B$-field-resolved Brody parameter $\eta(B)$ of the
theoretical coupled-channels data in Figs.~\ref{fig:distribution}(c)
and (d). A comparison with the experimental $\eta(B)$ shows excellent
agreement for $^{164}$Dy, while the agreement for $^{168}$Er is less
satisfactorily. A possible explanation for the discrepancies in $^{168}$Er
is the larger bin-to-bin fluctuations as bins contain fewer resonances
than for $^{164}$Dy.

For $^{164}$Dy the theoretical field-resolved Brody parameter in
Fig.~\ref{fig:distribution}(c) linearly increases from zero for small $B$
fields and saturates at $\eta(B)\approx 0.5$ for fields larger than 35
G where the size or width of the avoided crossings between weakly-bound
states is larger.  For $^{168}$Er in Fig.~\ref{fig:distribution}(d)
we find a much more rapid increase of $\eta(B)$ at small fields. This
is followed by a plateau at $\eta(B)\approx 0.5$ between $B=20$ G and
50 G, after which $\eta(B)\to0.9$ with an uncertainty of 0.2 close to
a Wigner distribution. The initial rise of $\eta(B)$ for both atomic
species is a consequence of weakly-bound vibrational levels, uncoupled
and randomly distributed when $B = 0$ G, that start to repel each other
as the Zeeman interaction increases in strength for increasing $B$. The
plateau at $\eta(B)\approx 0.5$ and the sudden increase of $\eta(B)$ to
one for $^{168}$Er have no simple explanation and are determined by the
not-fully-explored complex interplay between the Zeeman and anisotropic
inter-atomic interactions. It does, however, indicate that Wigner's
assumptions on ensembles of Hamiltonians do not hold for  fields below
50 G.


\section{Temperature dependence of the resonance density} \label{sec:Tdep}

\begin{figure*}[t]
\includegraphics[scale=0.29]{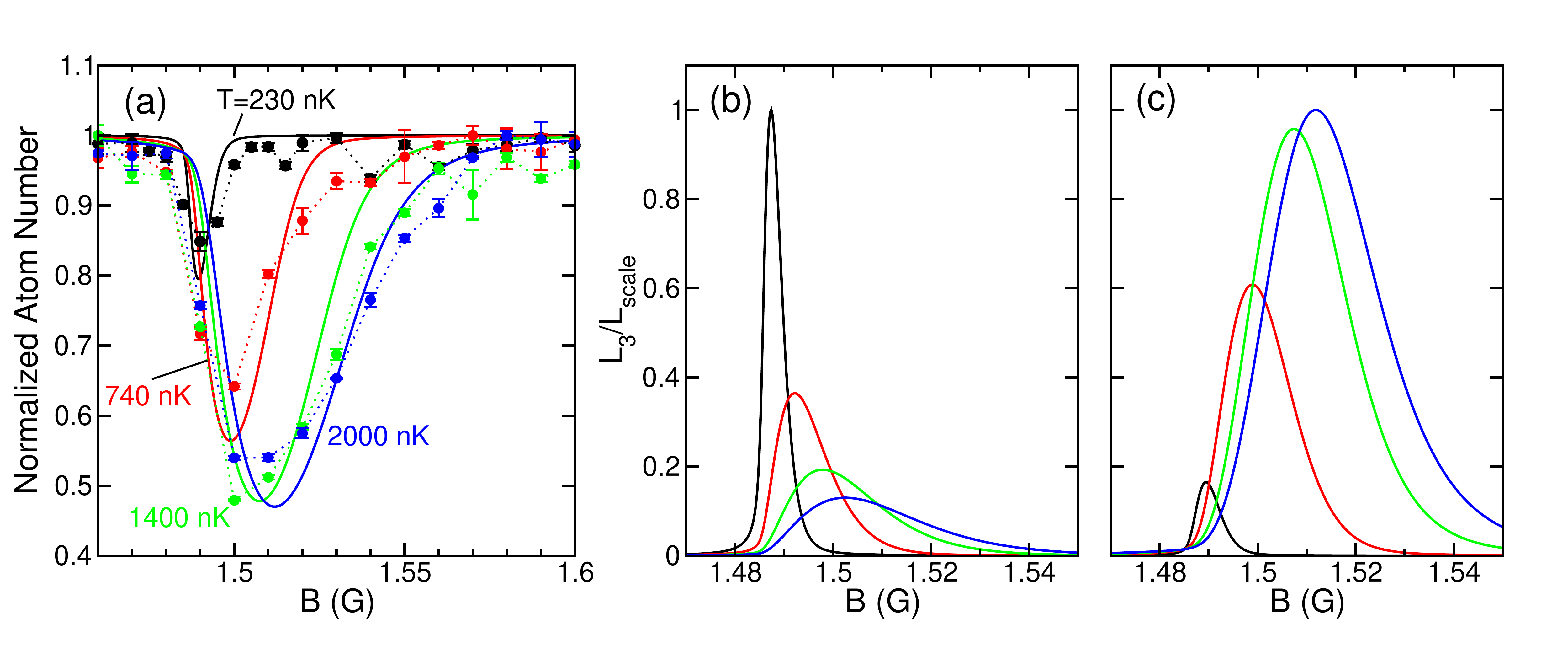}
\caption{(Color online) Line shapes for a strongly-temperature-dependent
$^{168}$Er Feshbach resonance near $B=1.48$ G.
Panel (a) shows experimental data (markers with error bars) as a function
of $B$ of the remaining atom number divided by the atom number away
from resonance measured 400 ms after initial preparation.  Black, red,
green, and blue markers correspond to data for temperatures $T=230$ nK,
740 nK, 1400 nK, and 2000 nK, respectively.  Dashed lines connecting the
markers guide the eye.  Solid lines are theoretical line shapes of the
remaining atom number based on the $d$-wave (${\cal N}=2$) recombination
rates shown in panel (c).  Panels (b) and (c) show simulated three-body
recombination rates for the same four temperatures assuming three-body
entrance $s$- (${\cal N}=0$) and $d$-wave  (${\cal N}=2$) scattering,
respectively.  Curves are based on a thermally-averaged line shape
discussed in the text.  Recombination rates are scaled such that the
largest value in each panel is one.  For both panels $\mu=3.1 \mu_B$
and $\Gamma_{\rm br}/k_B=250$ nK, while $\Gamma(E)/k_B=0.2(E/E_{\rm ref})^2$
nK in panel (b) and $\Gamma(E)/k_B=0.1(E/E_{\rm ref})^4$ nK in panel (c),
where $E_{\rm ref}/k_B=1000$ nK.  } 
\label{fig:Er_Tdep} 
\end{figure*}

We now describe the origin of the strong temperature dependence of some of
the resonances in our atom-loss spectra and thus explain the accompanying
increase of the resonance density.  Here, atom loss is solely due to
three-body recombination, where three ultra-cold atoms collide to form
a di-atomic molecule and an atom that both are lost from the atom trap.
Figure \ref{fig:Er_Tdep}(a) shows atom-loss spectra for one such resonance
for $^{168}$Er at four temperatures below 2 $\mu$K.  Atom loss, indeed,
is larger for larger temperatures, but we also observe a broadening of
the $B$-field width and a shift of the maximum loss position to larger $B$ fields.
Resonances with a weak  temperature dependence show
none of these behaviors. 

We show with an intuitive resonant ``trimer'' model that  a strongly
temperature-dependent resonance is due to scattering processes with
entrance $d$-wave channels even though the two-body $d$-wave centrifugal
barrier, $V_b/k_B=250$ $\mu$K, is a  hundred times  larger than our
highest temperature, where $k_B$ is the Boltzmann constant. The difference
in the power-law Wigner-threshold  behavior of the recombination rate
with collision energy for $s$- and $d$-wave entrance-channel collisions
can explain our observations.

Three-body recombination has been extensively  studied in the context of
Efimov physics \cite{Petrov2004,Kraemer2006,Braaten2006,Massignan2008}. We
follow Refs.~\cite{Suno2002,Wang2011,Wang2012} and start from a
coupled-channels description in the (mass scaled) hyperradius $\rho$,
which describes the size of the three-atomic system, and basis functions
in the five other hyperspherical coordinates that are $\rho$-dependent
eigen states of the squared ``grand-angular momentum operator''. Similar
to the coupled-channels description for two atoms, there are entrance,
open, and closed channels. The collision starts in one of the entrance
channels with atoms in the energetically-lowest Zeeman state and relative
three-body kinetic energy $E_3$, the dimer plus atom
are the open channels, and bound states in closed channels can
lead to resonances. These closed channels dissociate to three free-atom
states with at least one atom in a Zeeman level with higher internal
energy. The bound states are resonant ``trimer'' states giving
us our name for the model.  It should, however, be realized that their
origin lies in bound states of pairs of atoms and that the resonant
state is better thought of as a pair bound state that hops from pair to pair.
We define $E_3\equiv\hbar^3k_3^2/(2\mu_3)\equiv\mu_3 v_3^2/2$
with the three-body reduced mass $\mu_3=m/\sqrt{3}$, where  $k_3$ and $v_3$
are the  relative wave vector and velocity, respectively.

The potentials in the entrance channels have long-range repulsive
centrifugal potentials, governed by the asymptotic behavior of the
grand-angular momentum operator and depend on the relative orbital angular
momentum $\cal \vec  N$ of the three atoms.  In fact, the centrifugal
potentials are $ \hbar^2(\lambda+3/2)(\lambda+5/2)/(2\mu_3 \rho^2)$
with non-negative integer  quantum number $\lambda$ \cite{Suno2002}.
For ${\cal N}=0$ the least repulsive potential has $\lambda=0$, while
that for ${\cal N}=2$ has $\lambda=2$.

For an isolated trimer resonance  in a closed channel coupled to both
entrance and other open channels we can apply the resonance theories
by Fano and Feshbach and derive that the  recombination rate
coefficient at collision energy $E_3$ and entrance channel with
quantum number $\lambda$ is given by $L_3(E_3,B) = v_3\sigma(E_3,B)$,
where the  cross section $\sigma(E_3,B)=(2{\cal N}+1)192\pi^2
|S(E_3,B)|^2/k_3^5$ and
\[
 |S(E_3,B)|^2=  \frac{\Gamma(E_3) \Gamma_{\rm br}}{(E_3-\mu (B-B_0))^2+ (\Gamma_{\rm tot}(E_3)/2)^2}
\]
is a resonant expression for the square of a dimensionless $S$-matrix
element, where $B_0$ is the trimer resonance location and $\mu$ is
the magnetic moment of the resonant trimer relative to that of the
entrance channel.  The definition  for $|S(E_3,B)|^2$ also contains the
entrance-channel energy width $\Gamma(E_3)=A_{\lambda} E_3^{\lambda+2}$
to the trimer resonance with a characteristic power-law energy dependence
that reflects the threshold behavior of the scattering solutions in the
centrifugal potentials.  The energy width $\Gamma_{\rm br}$ determines the
decay or breakup rate of the resonance into the fast atom and dimer pair
and is independent of $E_3$. Finally,  $\Gamma_{\rm tot}(E_3)=\Gamma(E_3)+
\Gamma_{\rm br} $.  For simplicity we have assumed that non-resonant,
direct recombination from the entrance to open channels is weak.  We also
note that for ${\cal N}=0$ and $\lambda=0$, $L_3(E_3,B)$ approaches a
finite constant for $E_3\to0$ as expected.

In our experiments we have thermal samples of Er and we require
the thermally-averaged rate coefficient
\[ 
   L_3(T,B)=\frac {1}{Z}\int_0^\infty E^2 dE\, L_3(E,B)\, e^{-E/kT}
\]
and normalization $Z=\int_0^\infty E^2 dE \,e^{-E/kT}=2(kT)^3$.  In order to
increase signal to noise we have allowed a significant fraction of atoms
to be lost (See Fig.~\ref{fig:Er_Tdep}(c)), which assuming a homogeneous
sample can be modeled by the rate equation $dn(t)/dt=-3L_3(T,B) n^3(t)$
for atom density $n(t)$ \cite{Braaten2006} with solution
\[
   N(t_{\rm h},B)=\frac{N_{0}}{\sqrt{1+ 6L_3(T,B) n_0^2 t_{\rm h}}}\,,
\]
where $N(t_{\rm h},B)$ is the remaining atom number after hold time $t_h$,
$N_{0}$ is the initial atom number, and $n_0$ is the initial density.
This non-linear time evolution adds additional broadening to the lines.

Figures \ref{fig:Er_Tdep}(b) and (c) show our model event rates $L_3(T,B)$
as a function of $B$ for ${\cal N}=0$, $\lambda=0$ and ${\cal N}=2$,
$\lambda=2$ , respectively.  Curves are for the same four temperatures
as in Fig.~\ref{fig:Er_Tdep}(a).  A comparison of  panels (b) and (c)
shows a striking difference. The strongest features in panel (b) are for
the smallest temperatures, while those in panel (c) are for the largest
temperatures.  This behavior naturally follows from an approximation of
the integrant in $L_3(T,B)$ under the conditions $kT\gg \Gamma_{\rm
br}\gg\Gamma(E)$ \cite{Jones2000}.  In this limit the Lorentzian is
sharply peaked around $E_3=\mu (B-B_0)$ for $B>B_0$ and after some algebra
it follows that $L_3(T,B)$ as a function of $B$ has a maximum value
proportional to $(kT)^{\lambda-1}$ located at $B=B_0+(\lambda+2)kT/\mu$.
Consequently, for $\lambda=0$ and 2 the maximum loss rate decreases and
increases with $T$, respectively. Even for less restrictive parameter
values as used in Fig.~\ref{fig:Er_Tdep} this trend remains.

Our experimental data has a temperature trend as in panel (c). In fact,
Fig.~\ref{fig:Er_Tdep}(a) compares our experimental loss data with
model $N(t_{\rm h},B)$ for ${\cal N}=2$, $\lambda=2$  using the same
parameters as in Fig.~\ref{fig:Er_Tdep}(c) and requiring a $\approx 50\%$
maximum atom loss as in the experiment.  It is worth noting that, from our theoretical calculations, the magnetic field
width of $L_3(T,B)$ is noticeably smaller than that for $N(t_{\rm h},B)$
indicating that the finite hold time does indeed lead to broadening.
The agreement of the experimental data and the prediction of our model for the losses is satisfactory for all four temperatures
given the limitations and approximations within our modeling.  
We conclude that our strongly $T$-dependent resonances correspond to
$d$-wave or more precisely ${\cal N}=2$ entrance channel collisions. 
Note that we haven't observed any resonances with  temperature dependence similar to Fig.~\ref{fig:Er_Tdep}(b) in our spectra. In the case of resonances with three-body s-wave entrance channel, which would correspond to such a dependence, we infer that the loss spectra is saturated. This will be subject for future investigations.

As a corollary, this implies that for two colliding atoms as described
in Sec.~\ref{sec:micro} temperature-dependent resonances are due to
collisions with entrance $d$-waves for which there are multiple allowed
values of the total angular projection quantum number $M$.  Here M = -14 to -10
for bosonic Er and M = -18 to -14 for bosonic Dy. Numerical
computations, not presented here, show that their zero-field bound-states
and thus resonance locations are again uncorrelated and random.


\section{Conclusion}

In summary, we have experimentally and theoretically studied the resonant
scattering of ultracold Er and Dy atoms in a magnetic field.  We have
shown that chaotic scattering as witnessed by chaotic nearest-neighbor
spacings between Feshbach resonance locations emerges due to the
anisotropy in the molecular dispersion.

Our study has also revealed several unique features of colliding magnetic
lanthanides that have not been observed in any other ultracold atomic
system. These lanthanides are characterized by their exceptionally
large electron orbital angular momentum, which lead to large anisotropic
dispersion interactions between these atoms. Our theoretical estimate
shows that in both Er and Dy collisions the ratio of anisotropic to
isotropic dispersion interaction $\Delta C_6/C_6$ is about 10\%. This
anisotropy leads to significant splittings among the 48 and 81 {\it
gerade} short-range potentials that dissociate to the ground-state
atomic limits of Er and Dy, respectively. We have shown that each potential
has its own ro-vibrational structure, which by coriolis forces and
the Zeeman interaction interacts with that of other potentials, creating a dense 
distribution of levels near the threshold and initiating chaos. In fact, we find a 
very large number of partial waves contributing to the creation of Fano-Feshbach resonances.

On the other hand, if we just consider the anisotropy from the magnetic
dipole-dipole interaction alone, our coupled-channel calculations indicate
that chaos in the level distribution does not appear. The strength of
the dipole-dipole interaction is too small.  In addition, we have shown
that the NNS distributions for Dy and Er are very similar, as can be
expected from their similar $\Delta C_6/C_6$ ratio. The difference in
their magnetic moment only plays a small role. This further confirms
that chaos is due to the anisotropic dispersion interaction.

The distribution of Feshbach resonances of ultra-cold ground-state
alkali-metal, alkaline-earth, Yb, and Cr atoms, as experimental studies
have shown, are not chaotic. This is because these atoms have a zero
electron orbital angular momentum and, hence, only an isotropic dispersion
interaction. Eventhough alkali-metal and Cr atoms have a non-zero magnetic
moment of 1$\mu_B$ and 6$\mu_B$, respectively, these moments 
do not lead to chaos. We would expect that other magnetic lanthanides and
actinides with non-zero orbital angular momentum will exhibit chaotic properties
in their collisions. In addition, collisions between mixed species,
such as magnetic lanthanides and alkali-metals, like K+Dy or Na+Er,
might be susceptible to chaos. A first theoretical analysis for Li+Er
\cite{Gonzalez2015}, however, estimates a small 2\% dispersion anisotropy
and no chaos is predicted.

Another interesting property of magnetic lanthanide gases is the
extreme sensitivity of the atom-loss spectra and, in essence, three-body
recombination to the temperature.  This phenomenon was first observed
in Ref.~\cite{DyFeshbach} for loss spectra of Dy. The number of Dy
resonances increases by 50\% when the temperature is increased from
420 nK to 800 nK.  Here, we observe a 25\% increase in the Er resonance
density when the temperature rises from 250 nK to 1400 nK. We have shown
by a comparison of resonance profiles taken at several temperatures and
predictions of a theoretical model of three-body recombination via the
formation of a trimer, or more precisely of a shared pair bound state,
that the origin of the temperature-dependent resonances lies in the
``partial wave'' of the three-atom entrance channel.  Entrance channels with
zero and non-zero total orbital angular momentum $\cal N$ lead to line
shapes with a different temperature behavior. Those with ${\cal N}=0$ or
``$s$-wave'' entrance channels have sharply decreasing recombination rates
with temperature, whereas those with ${\cal N}=2$ or ``$d$-wave'' entrance
channels have an increasing recombination rate.  Temperature sensitive
resonances can only be explained by ``$d$-wave'' collisions.  It is worth
noting that for alkali-metal-atom collisions a number of entrance-channel $p$-wave
resonances have been observed (See for example Ref.~\cite{Chin2004}
for cesium). Analysis of the temperature dependent rate coefficient, however,
was not performed.

\section*{Acknowledgments}

The Stuttgart group thanks Axel Griesmaier for the support at the early
stage of the experiment.  The Dy work is supported by the German Research
Foundation (DFG) within SFB/TRR21. H.K. acknowledges support by the
``Studienstiftung des deutschen Volkes''. The Er work is supported by the
Austrian Ministry of Science and Research (BMWF) and the Austrian Science
Fund (FWF) through a START grant under Project No. Y479-N20 and by the
European Research Council under Project No. 259435. K. A. was supported
within the Lise-Meitner program of the FWF. 
L.C. acknowledges support by the FWF through SFB FoQuS.
Work at Temple University
is supported by the AFOSR grant No. FA9550-14-1-0321 and the NSF grant
No. PHY-1308573. The work at JQI is supported by the NSF grant No. PHY-1506343.

\bibliography{DyEr}

\end{document}